\documentclass[twocolumn,nofootinbib,amsmath,amssymb,a4paper]{revtex4}

\usepackage{graphicx}
\usepackage{dcolumn}
\usepackage{bm}
\usepackage[dvips]{color}
\usepackage{epsfig}
\usepackage{amssymb,amsmath}
\usepackage{lscape}

\begin{document}

\title{On model-independent measurement of the angle \boldmath{$\phi_3$} 
using Dalitz plot analysis.}

\author{Alex Bondar and Anton Poluektov}
\affiliation{Budker Institute of Nuclear Physics, Novosibirsk, Russia}
\email{A.O.Poluektov@inp.nsk.su}

\newcommand{\bdk}{$B^{\pm}\to DK^{\pm}$}
\newcommand{\bdtk}{$B^{\pm}\to \tilde{D}K^{\pm}$}

\newcommand{\bdsk}{$B^{\pm}\to D^{*}K^{\pm}$}
\newcommand{\bdstk}{$B^{\pm}\to \tilde{D}^{*}K^{\pm}$}

\newcommand{\bdgk}{$B^{\pm}\to D^{*}(D\gamma)K^{\pm}$}
\newcommand{\bdgtk}{$B^{\pm}\to \tilde{D}^{*}(D\gamma)K^{\pm}$}

\newcommand{\bdks}{$B^{\pm}\to DK^{*\pm}$}
\newcommand{\bdtks}{$B^{\pm}\to \tilde{D}K^{*\pm}$}
\newcommand{\bdksnr}{$B^{\pm}\to DK^0_S\pi^{\pm}$}

\newcommand{\bddsk}{$B^{\pm}\to D^{(*)}K^{\pm}$}
\newcommand{\bddstk}{$B^{\pm}\to \tilde{D}^{(*)}K^{\pm}$}

\newcommand{\bddsks}{$B^{\pm}\to D^{(*)}K^{(*)\pm}$}
\newcommand{\bddstks}{$B^{\pm}\to \tilde{D}^{(*)}K^{(*)\pm}$}

\newcommand{\bdpi}{$B^{\pm}\to D\pi^{\pm}$}
\newcommand{\bdtpi}{$B^{\pm}\to \tilde{D}\pi^{\pm}$}

\newcommand{\bdspi}{$B^{\pm}\to D^{*}\pi^{\pm}$}
\newcommand{\bdstpi}{$B^{\pm}\to \tilde{D}^{*}\pi^{\pm}$}

\newcommand{\bdgpi}{$B^{\pm}\to D^{*}(D\gamma)\pi^{\pm}$}
\newcommand{\bdgtpi}{$B^{\pm}\to \tilde{D}^{*}(D\gamma)\pi^{\pm}$}

\newcommand{\bddspi}{$B^{\pm}\to D^{(*)}\pi^{\pm}$}
\newcommand{\bddstpi}{$B^{\pm}\to \tilde{D}^{(*)}\pi^{\pm}$}

\newcommand{\dsdpi}{$D^{*\pm}\to D\pi^{\pm}$}
\newcommand{\dsdpis}{$D^{*\pm}\to D\pi_s^{\pm}$}

\newcommand{\dkpp}{$\overline{D}{}^0\to K^0_S\pi^+\pi^-$}
\newcommand{\dtkpp}{$\tilde{D}\to K^0_S\pi^+\pi^-$}

\newcommand{\dn}{$D^0$}
\newcommand{\dnbar}{$\overline{D}{}^0$}

\begin{abstract}
  This report shows the latest results on the study of the method to determine 
  the angle $\phi_3$ of the unitarity triangle using Dalitz plot analysis of 
  $D^0$ decay from \bdk\ process in a model-independent way. We concentrate 
  on the case with a limited charm data sample, which will be available from 
  the CLEO-c collaboration in the nearest future, with the main goal to 
  find the optimal strategy for $\phi_3$ extraction. 
\end{abstract}

\maketitle

\section{Introduction}

The measurement of the angle $\phi_3$ ($\gamma$) of the unitarity 
triangle using Dalitz plot analysis of the $D^0\to K^0_S\pi^+\pi^-$ decay 
from $B^{\pm}\to DK^{\pm}$ process, introduced by Giri {\em et al.}~\cite{giri} and Belle 
collaboration~\cite{binp_belle} and successfully implemented by 
BaBar~\cite{babar_phi3} and Belle~\cite{belle_phi3}, presently 
offers the best constraints on this quantity. However, this technique is 
sensitive to the choice of the model used to describe the three-body $D^0$ 
decay. Currently, this uncertainty is estimated to be 
$\sim 10^{\circ}$ and due to large statistical error does not affect the 
precision of $\phi_3$ measurement. As the amount of $B$ factory data 
increases, though, this uncertainty will become a major limitation. 
Fortunately, a model-independent approach exists (see~\cite{giri}), which 
uses the data of the $\tau$-charm factory to obtain the missing information 
about the $D^0$ decay amplitude. 

In our previous study of the model-independent Dalitz analysis 
technique~\cite{phi3_modind} we have implemented a procedure proposed 
by Giri {\em et al.} involving the division of the Dalitz plots into bins, 
and shown that this procedure allows to measure the phase $\phi_3$ with 
the statistical precision only 30--40\% worse than in the unbinned 
model-dependent case. We did not attempt to optimize the binning and 
mainly considered a high-statistics limit with an aim to estimate the 
sensitivity of the future super-B factory. 

The data useful for model-independent measurement are presently available 
from the CLEO-c experiment~\cite{cleoc}. 
CLEO-c collected an integrated luminosity of 
280 pb$^{-1}$ at the $\psi(3770)$ resonance decaying to $D\bar{D}$. By 
the end of CLEO-c operation this statistics will grow up to 
750~fb$^{-1}$~\cite{david}. 
This corresponds to $\sim 1000$ events where $D$ meson in a $CP$ eigenstate
decays to $K^0_S\pi^+\pi^-$, and twice as much events of 
$\psi(3770)\to D^0\overline{D}{}^0$ with both $D$ mesons decaying to 
$K^0\pi^+\pi^-$. Both of these processes include the information useful 
for a model-independent $\phi_3$ measurement. 
In this paper, we report on studies of the model-independent 
approach with a limited statistics of both $\psi(3770)$ and $B$ data, 
using both $D_{CP}\to K^0_S\pi^+\pi^-$ and 
$(K^0\pi^+\pi^-)_D(K^0\pi^+\pi^-)_D$ final states.

\section{Model-independent approach}

The density of \dkpp\ Dalitz plot is given by the absolute value of 
the amplitude $f_D$ squared: 
\begin{equation}
  p_D=p_D(m^2_+, m^2_-)=|f_D(m^2_+, m^2_-)|^2
  \label{p_d}
\end{equation}
In the case of no $CP$-violation in $D$ decay the density of the \dn\ 
decay $\bar{p}_D$ equals to
\begin{equation}
  \bar{p}_D=|\bar{f}_D|^2=p_D(m^2_-, m^2_+). 
\end{equation}
Then the density of the $D$ decay Dalitz plot from \bdk\ process is 
expressed as
\begin{equation}
\begin{split}
  p_{B^{\pm}}=&|f_D+ r_Be^{i(\delta_B\pm\phi_3)}\bar{f}_D|^2=\\
              &p_D+r_B^2\bar{p}_D+2\sqrt{p_D\bar{p}_D}(x_{\pm}c+y_{\pm}s), 
  \label{p_b}
\end{split}
\end{equation}
where $x_{\pm},y_{\pm}$ include the value of $\phi_3$ and other related 
quantities, the strong phase $\delta_B$ of the \bdk\ decay, and amplitude 
ratio $r_B$:
\begin{equation}
  x_{\pm}=r_B\cos(\delta_B\pm\phi_3); \;\;\;
  y_{\pm}=r_B\sin(\delta_B\pm\phi_3). 
\end{equation}
The functions $c$ and $s$ are the cosine and sine of the strong phase 
difference $\Delta\delta_D$ between the symmetric Dalitz plot points: 
\begin{equation}
\begin{split}
  c=&\cos(\delta_D(m^2_+,m^2_-)-\delta_D(m^2_-,m^2_+))=\cos\Delta\delta_D; \\
  s=&\sin(\delta_D(m^2_+,m^2_-)-\delta_D(m^2_-,m^2_+))=\sin\Delta\delta_D. 
\end{split}
\end{equation}
The phase difference $\Delta\delta_D$ can be obtained from the sample of 
$D$ mesons in a $CP$-eigenstate, decaying to $K^0_S\pi^+\pi^-$. The Dalitz plot 
density of such decay is 
\begin{equation}
  p_{CP}=|f_D\pm \bar{f}_D|^2=p_D+\bar{p}_D\pm 2\sqrt{p_D\bar{p}_D}c
  \label{p_cp}
\end{equation}
(the normalization is arbitrary). 
Decays of $D$ mesons in $CP$ eigenstate to $K^0_S\pi^+\pi^-$ can be obtained 
in the process, {\it e.g.} $e^+e^-\to\psi(3770)\to D\bar{D}$, 
where the other (tag-side) $D$ meson is reconstructed in the $CP$ eigenstate, 
such as $K^+K^-$ or $K^0_S\omega$. 

Another possibility is to use a sample, where both $D$ mesons (we denote them 
as $D$ and $D'$) from the $\psi(3770)$ meson decay into the 
$K^0\pi^+\pi^-$ state~\cite{kppkpp}. 
Since $\psi(3770)$ is a vector, two $D$ mesons are produced in a $P$-wave, 
and the wave function of the two mesons is antisymmetric. Then the
four-dimensional density of two correlated Dalitz plots is
\begin{equation}
\begin{split}
  p_{\rm corr}&(m_+^2,m_-^2,m'^2_+,m'^2_-)=|f_D\bar{f}_D'-f_D'\bar{f}_D|^2=\\
     &p_D\bar{p}_D'+\bar{p}_Dp_D'-2\sqrt{p_D\bar{p}_Dp_D'\bar{p}_D'}(cc'+ss'),
  \label{p_corr}
\end{split}
\end{equation}
This decay is sensitive to both $c$ and $s$ for the price of having to deal with 
the four-dimensional phase space. 

In a real experiment, one measures scattered data rather than a 
probability density. Two options of dealing with real data are possible: 
a binned approach, or a scatter plot smoothing using nonparametric 
density estimation. The latter could be useful to reach the statistical 
sensitivity equivalent to the model-dependent case. However, in this 
paper we show that using the appropriate binning this is also possible. 

\section{Binned analysis with $D_{CP}$ data}

The binned approach was proposed by Giri {\em et al.} \cite{giri}. 
Assume that the Dalitz plot is divided into $2\mathcal{N}$ bins symmetrically
to the exchange $m^2_-\leftrightarrow m^2_+$. The bins are denoted
by the index $i$ ranging from $-\mathcal{N}$ to $\mathcal{N}$ (excluding 0); 
the exchange $m^2_+ \leftrightarrow m^2_-$ corresponds to the exchange 
$i\leftrightarrow -i$. Then the expected number of events 
in the bins of the Dalitz plot of $D$ decay from \bdk\ is 
\begin{equation}
  \langle N_i\rangle = h_B[K_i + r_B^2K_{-i} + 2\sqrt{K_iK_{-i}}(xc_i+ys_i)], 
  \label{n_b}
\end{equation}
where $K_i$ is the number of events in the bins in the Dalitz plot 
of the $D^0$ in a flavor eigenstate, $h_B$ is the normalization
constant. Coefficients $c_i$ and $s_i$, which include the information about 
the cosine and sine of the phase difference, are given by
\begin{equation}
  c_i=\frac{\int\limits_{\mathcal{D}_i}
            \sqrt{p_D\bar{p}_D}
            \cos(\Delta\delta_D(m^2_+,m^2_-))d\mathcal{D}
            }{\sqrt{
            \int\limits_{\mathcal{D}_i}p_Dd\mathcal{D}
            \int\limits_{\mathcal{D}_i}\bar{p}_Dd\mathcal{D}
            }}, 
  \label{cs}
\end{equation}
$s_i$ is defined similarly with cosine substituted by sine. 
Here $\mathcal{D}_i$ is the bin region, over which the integration is 
performed. Note that $c_i=c_{-i}$, $s_i=-s_{-i}$ and $c_i^2+s_i^2\leq 1$
(the equality $c_i^2+s_i^2=1$ being satisfied if the amplitude is constant
across the bin). 

The coefficients $K_i$ are obtained precisely from a very large sample 
of $D^0$ decays in the flavor eigenstate, which is accessible at $B$-factories. 
The expected number of events in the Dalitz plot of $D_{CP}$ decay equals to
\begin{equation}
  \langle M_i\rangle = h_{CP}[K_i + K_{-i} + 2\sqrt{K_iK_{-i}}c_i], 
\end{equation}
and thus can be used to obtain the coefficient $c_i$. As soon as the $c_i$ and 
$s_i$ coefficients are known, one can obtain $x$ and $y$ values (hence, $\phi_3$
and other related quantities) by a maximum likelihood fit using equation (\ref{n_b}). 

Note that now the quantities of interest $x$ and $y$ (and consequently $\phi_3$)
have two statistical errors: one due to a finite sample of \bdk\ data, and 
due to $D_{CP}\to K^0_S\pi^+\pi^-$ statistics. We will refer to these errors 
as $B$-statistical and $D_{CP}$-statistical, respectively. 

Obtaining $s_i$ is a major problem in this analysis. If the binning is fine 
enough, so that both the phase difference and the amplitude remain 
constant across the area of each bin, expressions (\ref{cs}) reduce to 
$c_i=\cos(\Delta\delta_D)$ and $s_i=\sin(\Delta\delta_D)$, so $s_i$ can 
be obtained as $s_i=\pm\sqrt{1-c_i^2}$. Using this equality if the amplitude
varies will lead to the bias in the $x,y$ fit result. Since $c_i$ is obtained 
directly, and $s_i$ is overestimated by the absolute value, the bias will 
mainly affect $y$ determination, resulting in lower absolute values of $y$. 

Our studies \cite{phi3_modind} show that the use of equality $c_i^2+s_i^2=1$ 
is satisfactory for the number of bins around 200 or more, which cannot be used 
with presently available $D_{CP}$ data. It is therefore essential to find a 
relatively coarse binning (the number of bins being 10--20) which 
a) allows to extract $s_i$ from $c_i$ with low bias, and 
b) has the sensitivity to the $\phi_3$ phase 
comparable to the unbinned model-dependent case. 

Fortunately, both the a) and b) requirements appear to be equivalent. 
To determine the $B$-statistical sensitivity of a certain binning, 
let's define a quantity $Q$ --- a ratio of a statistical sensitivity 
to that in the unbinned case. Specifically, $Q$ relates the number of 
standard deviations by which the number of events in bins is changed by 
varying parameters $x$ and $y$, to the number of standard deviations 
if the Dalitz plot is divided into infinitely small regions 
(the unbinned case):
\begin{equation}
  Q^2=\frac{\sum\limits_i
    \left(\frac{1}{\sqrt{N_i}}\frac{dN_i}{dx}\right)^2+
    \left(\frac{1}{\sqrt{N_i}}\frac{dN_i}{dy}\right)^2
  }{
    \int\limits_{\mathcal{D}}\left[
      \left(\frac{1}{\sqrt{|f_B|^2}}\frac{d|f_B|^2}{dx}\right)^2+
      \left(\frac{1}{\sqrt{|f_B|^2}}\frac{d|f_B|^2}{dy}\right)^2
    \right]\,d\mathcal{D}
  }, 
\end{equation}
where $f_B=f_D+(x+iy)\bar{f}_D$, $N_i=\int_{\mathcal{D}_i}|f_B|^2d\mathcal{D}$. 

Since the precision of $x$ and $y$ weakly depends on the values
of $x$ and $y$ \cite{phi3_modind}, we can take for simplicity $x=y=0$. 
In this case one can show that
\begin{equation}
  Q^2|_{x=y=0}=\sum\limits_i(c^2_i+s^2_i) N_i\left/\sum\limits_i N_i\right. 
\end{equation}
Therefore, the binning which satisfies $c_i^2+s^2_i=1$ ({\em i.e.}
the absence of bias if $s_i$ is calculated as $\sqrt{1-c_i^2}$) also 
has the same sensitivity as the unbinned approach. The factor $Q$ defined 
this way is not necessarily the best measure of the 
binning quality (the binning with higher $Q$ can be insensitive 
to either $x$ or $y$, which is impractical from the point of measuring 
$\phi_3$), but it allows an easy calculation and correctly reproduces the 
relative quality for a number of binnings we tried in our simulation. 

The choice of the optimal binning naturally depends on the $D^0$ model. 
In our studies we use the two-body amplitude obtained in the latest Belle 
$\phi_3$ Dalitz analysis \cite{belle_phi3}. 

\begin{figure}
  \epsfig{file=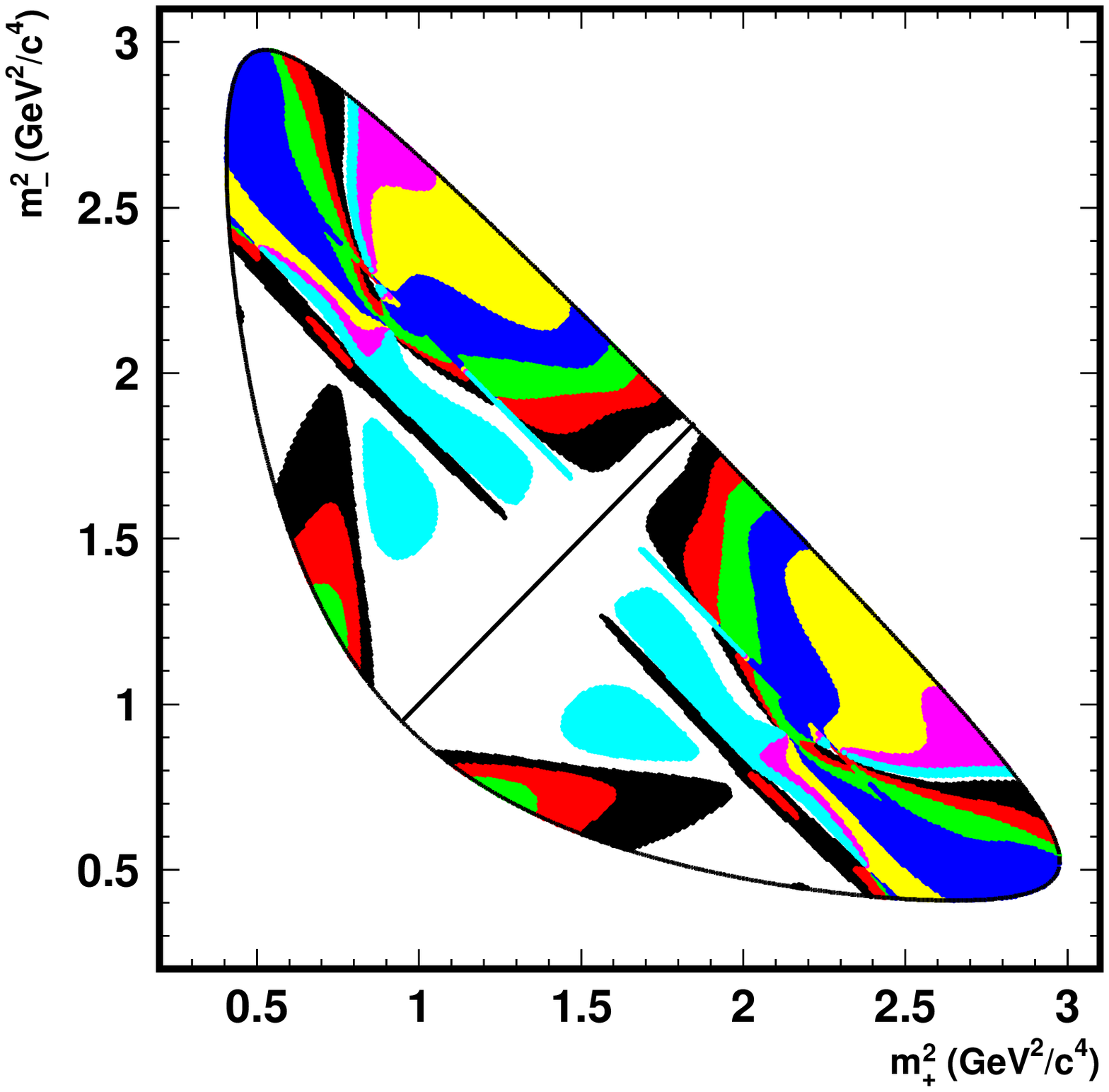, width=0.23\textwidth}
  \epsfig{file=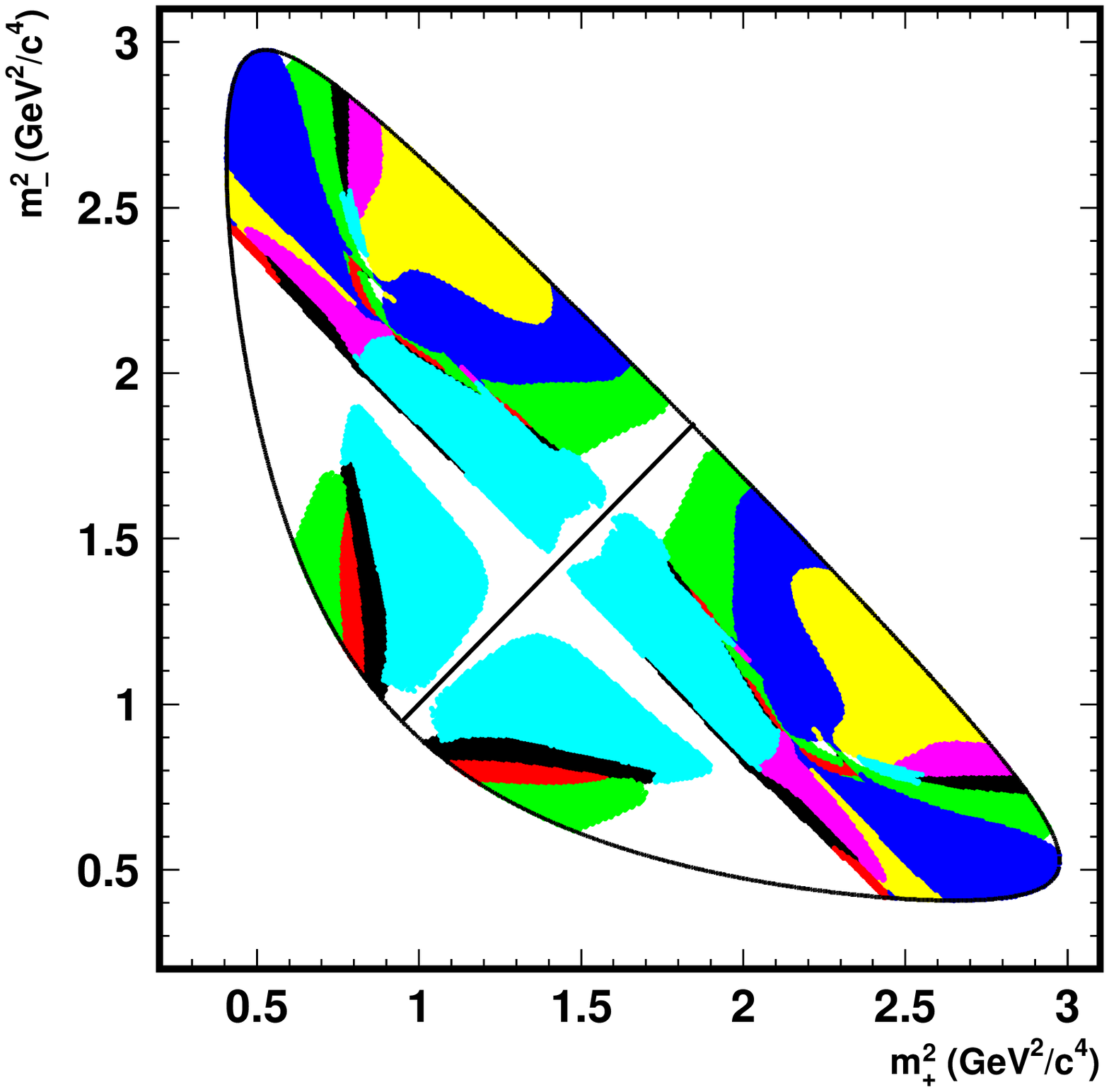, width=0.23\textwidth}
  \vspace{-\baselineskip}
  \caption{Divisions of the \dkpp\ Dalitz plot. Uniform binning of 
           $\Delta\delta_D$ strong phase difference with $\mathcal{N}=8$ (left), 
           and the binning obtained by variation of the latter to maximize 
           the sensitivity factor $Q$ (right). 
           }
  \label{binning}
\end{figure}

From the consideration above it is clear that a good approximation 
to the optimal binning is the one obtained from 
the uniform division of the strong phase difference $\Delta\delta_D$. 
In the half of the Dalitz plot $m^2_+<m^2_-$ ({\it i.e.} the bin index $i>0$) 
the bin $\mathcal{D}_i$ is defined by condition
\begin{equation}
   2\pi(i-1/2)/\mathcal{N}<\Delta\delta_D(m^2_+, m^2_-)<
   2\pi(i+1/2)/\mathcal{N}, 
\end{equation}
in the remaining part ($i<0$) the bins are defined symmetrically. 
We will refer to this binning as $\Delta\delta_D$-binning. 
As an example, such a binning with $\mathcal{N}=8$ is shown in 
Fig.~\ref{binning}~(a). Although the phase difference variation across the 
bin is small by definition, the absolute value of the amplitude can 
vary significantly, so the condition $c^2_i+s^2_i=1$ is not satisfied 
exactly. The values of $c_i$ and $s_i$ in this binning are shown in 
Fig.~\ref{cspic} with crosses. 

Figure~\ref{binning}~(b) shows the division with $\mathcal{N}=8$ obtained by 
continuous variation of the $\Delta\delta_D$-binning to maximize the factor $Q$. 
The sensitivity factor $Q$ increases to 0.89 compared to 0.79 for 
$\Delta\delta_D$-binning. 

\begin{table*}
  \caption{Statistical precision of $(x,y)$ determination using 
           different binnings and with an unbinned approach. The errors
           correspond to 1000 events in both the $B$ and 
           $D_{CP}$ ($(K^0\pi\pi)^2$) samples. }
  \label{stat}
  \begin{tabular}{|l||c||c|c||c|c||c|c|} 
    \hline
                                &   & \multicolumn{2}{|c||}{$B$-stat. err.} & 
                                      \multicolumn{2}{|c||}{$D_{CP}$-stat. err.} & 
                                      \multicolumn{2}{|c|}{$(K^0\pi\pi)^2$-stat.
                                      err.} \\ 
                                      \cline{3-8}
    Binning                     & Q & $\sigma_{x}$ & $\sigma_{y}$ & $\sigma_{x}$ & $\sigma_{y}$ & $\sigma_{x}$ & $\sigma_{y}$ \\ 
    \hline
    $\mathcal{N}=8$ (uniform)            & 0.57 & 0.0331 & 0.0600 & 0.0053 & 0.0097 & 0.0145 & 0.0322 \\
    $\mathcal{N}=8$ ($\Delta\delta_D$)   & 0.79 & 0.0273 & 0.0370 & 0.0042 & 0.0072 & 0.0050 & 0.0095 \\
    $\mathcal{N}=8$ (optimal)            & 0.89 & 0.0232 & 0.0324 & 0.0058 & 0.0114 & 0.0082 & 0.0114 \\
    $\mathcal{N}=19$ (uniform)           & 0.69 & 0.0274 & 0.0549 & 0.0042 & 0.0112 &      - &      - \\
    $\mathcal{N}=20$ ($\Delta\delta_D$)  & 0.82 & 0.0266 & 0.0350 & 0.0048 & 0.0074 &      - &      - \\
    $\mathcal{N}=20$ (optimal)           & 0.96 & 0.0223 & 0.0290 & 0.0078 & 0.0110 &      - &      - \\
    \hline
    Unbinned                    & -    & 0.0213 & 0.0279 & -      & -      &      - &      - \\ 
    \hline
  \end{tabular}
\end{table*}

We perform a toy MC simulation to study the statistical sensitivity of the 
different binning options. We use the amplitude from the Belle 
analysis~\cite{belle_phi3}
to generate decays of flavor $D^0$, $D_{CP}$, and $D$ from \bdk\ decay
to the $K^0_S\pi^+\pi^-$ final state according to the probability density given by 
(\ref{p_d}), (\ref{p_cp}) and (\ref{p_b}), respectively. 
To obtain the $B$-statistical error we use a large 
number of $D^0$ and $D_{CP}$ decays, while the generated number of $D$ decays 
from the \bdk\ process ranges from 100 to 100000. For each number of $B$ decay 
events, 100 samples are generated, and the statistical errors of 
$x$ and $y$ are obtained from the spread of the fitted values. 
A study of the error due to $D_{CP}$ statistics is performed similarly, 
with a large number of $B$ decays, and the statistics of $D_{CP}$ decays
varied. Both errors are checked to satisfy the square root scaling. 

The binning options used are $\Delta\delta_D$-binning with $\mathcal{N}=8$ 
and $\mathcal{N}=20$, 
as well as ``optimal" binnings with maximized $Q$ obtained from these two
with a smooth variation of the bin shape. 
Note that the ``optimal" binning with $\mathcal{N}=20$ offers the 
$B$-statistical sensitivity only 4\% worse than an unbinned technique. 
For comparison, we use the binnings with the uniform division into rectangular
bins (with $\mathcal{N}=8$ and $\mathcal{N}=19$ in the allowed phase space, 
the ones which are denoted as 3x3 and 5x5 in~\cite{phi3_modind}). 

The $B$- and $D_{CP}$-statistical precision of different binning options, 
recalculated to 1000 events of both $B$ and $D_{CP}$ samples, 
as well as their calculated values of the factor $Q$, 
are shown in Table~\ref{stat}. 
In the present study we use the errors of parameters $x$ and $y$ 
rather than $\phi_3$ as a measure of the statistical power since they are 
nearly independent of the actual values of $\phi_3$, strong phase $\delta$ and 
amplitude ratio $r_B$. The error of $\phi_3$ can be obtained from these
numbers given the value of $r_B$. The factor $Q$ reproduces the 
ratio of the values $\sqrt{1/\sigma_x^2+1/\sigma_y^2}$ for the binned 
and unbinned approaches with the precision of 1--2\%. While the binning with 
maximized $Q$ offers better $B$-statistical 
sensitivity, the best $D_{CP}$-statistical precision 
of the options we have studied is reached for the 
$\Delta\delta_D$-binning. However, for the expected amount of experimental 
data of $B$ and $D_{CP}$ decays the $B$-statistical error dominates, 
therefore, slightly worse precision due to $D_{CP}$ statistics does 
not affect significantly the total precision. 

Using $\Delta\delta_D$-binning, the following combination of the binned 
and unbinned approaches is possible, which allows to reach $B$-statistical
precision equivalent to the unbinned case. Assume the number of 
$\Delta\delta_D$-bins
is large enough, so $\cos\Delta\delta_D$ and $\sin\Delta\delta_D$ remain 
almost constant across the bin area. At the level of current precision
this is reached already for a number of bins as small as 10--20. Then
\begin{equation}
  c_i=\cos\Delta\delta_D\frac{\int\limits_{\mathcal{D}_i}
            \sqrt{p_D\bar{p}_D}d\mathcal{D}
            }{\sqrt{
            \int\limits_{\mathcal{D}_i}p_Dd\mathcal{D}
            \int\limits_{\mathcal{D}_i}\bar{p}_Dd\mathcal{D}
            }}, 
\end{equation}
where the integrals can be calculated from the flavor $D$ data sample. 
Therefore, it is possible to obtain $\cos\Delta\delta_D$ from 
$c_i$ (and consequently $\sin\Delta\delta_D$), and use them in 
the expression for the probability density (\ref{p_b}) to perform
the unbinned $B$-data fit thus obtaining the best possible $B$-statistical 
precision.

We have considered the choice of the optimal binning only from the 
point of statistical power. However, the conditions to satisfy low model 
dependence are quite different. Since the bins in the binning options 
we have considered are sufficiently large, the requirement that the phase 
does not change over the bin area is a strong model assumption. 
We have performed toy MC simulation to study the model dependence. 
While the binning was kept the same as in the statistical power study
(based on the phase difference from the default $D^0$ amplitude), 
the amplitude used to generate $D^0$, $D_{CP}$ and \bdk\ decays
was altered in the same way as in the Belle study of the model-dependence 
in the unbinned analysis~\cite{belle_phi3}. 
As a result, the same bias of $\Delta\phi_3\sim 10^{\circ}$ is observed
as in unbinned analysis. We remind that the cause of this bias is a fixed 
relation between the $c_i$ and $s_i$. Therefore, proposed binning options, 
although providing good statistical precision, are not flexible enough
to provide also a low model dependence. To minimizie the model dependence, 
the bin size should be kept as small as possible, 
therefore, uniform binning is more preferred. 

In a real analysis, one can control the model error by testing 
if the amplitude used to define binning is compatible with the observed 
$D_{CP}$ data. This can be done, {\em e.g.,} by dividing each bin and 
comparing calculated values of $c_i$ in its parts, or by comparing the 
expected and observed numbers of events in each bin. 

We conclude that the method of $\phi_3$ determination
using only $D_{CP}$ data is only asymptotically 
model-independent, since for any finite bin size the calculation 
of $s_i$ is done using model assumptions of the $\Delta\delta_D$
variations across the bin. Increasing the $D_{CP}$
data set, however, allows to apply a finer binning and therefore 
reduce the model error due to the variation of the phase difference. 

\section{Binned analysis with correlated $D^0\to K^0\pi\pi$ data}

The use of the $\psi(3770)$ decays where both neutral $D$ mesons decay to the
$K^0_S\pi^+\pi^-$ state allows to significantly increase the amount of data
useful to extract phase information in $D^0$ decay. It is also possible 
to detect events of $\psi(3770)\to (K^0_S\pi^+\pi^-)_D (K^0_L\pi^+\pi^-)_D$, 
where $K^0_L$ is not reconstructed, and its momentum is obtained from kinematic 
constraints. The number of these events is approximately twice that of 
$(K^0_S\pi\pi)^2$ due to combinatorics. We will refer 
to both of these processes as $(K^0\pi\pi)^2$. 

In the case of a binned analysis, the number of events in the region of the
$(K^0_S\pi\pi)^2$ phase space is 
\begin{equation}
\begin{split}
  \langle M\rangle_{ij} = h_{\rm corr}[&K_i K_{-j} + K_{-i} K_j - \\
    &2\sqrt{K_iK_{-i}K_jK_{-j}}(c_i c_j + s_i s_j)]. 
\end{split}
\end{equation}
In the case of the $(K^0_S\pi^+\pi^-)(K^0_L\pi^+\pi^-)$ final state, 
the interference term changes its sign. 
Here two indices correspond to two $D$ mesons from $\psi(3770)$ decay. 
It is logical to use the same binning as in the case of $D_{CP}$
statistics to improve the precision of the determination of 
$c_i$ coefficients, and to obtain $s_i$ from data without model 
assumptions, contrary to $D_{CP}$ case. The obvious advantage 
of this approach is its being unbiased for any finite 
$(K^0\pi\pi)^2$ statistics (not asymptotically as in the case of $D_{CP}$ data).

Note that in contrast to $D_{CP}$ analysis, where the sign of $s_i$
in each bin was undetermined and has to be fixed using model assumptions, 
$(K^0_S\pi^+\pi^-)$ analysis has only a four-fold ambiguity: change of 
the sign of all $c_i$ or all $s_i$. In combination with $D_{CP}$ analysis, 
where the sign of $c_i$ is fixed, this ambiguity reduces to only two-fold.  
One of the two solutions can be chosen based on a weak model assumption
(incorrect $s_i$ sign corresponds to complex-conjugated $D$ decay
amplitude, which violates causality requirement when parameterized with 
the Breit-Wigner amplitudes). 

Coefficients $c_i$, $s_i$ can be obtained by minimizing the negative 
logarithmic likelihood function
\begin{equation}
  -2\log\mathcal{L}=-2\sum\limits_{i,j}\log P(M_{ij}, \langle M\rangle_{ij}), 
  \label{corr_lh}
\end{equation}
where $P(M,\langle M\rangle)$ is the Poisson probability to get $M$ events with the 
expected number of $\langle M\rangle$ events. 

The number of bins in the 4-dimensional phase space is $4\mathcal{N}^2$
rather than $2\mathcal{N}$ in the $D_{CP}$ case. Since the expected number of 
events in correlated $K^0_S\pi\pi$ data is of the same order as for $D_{CP}$, 
the bins will be much less populated. This, however, does not affect the 
precision of $c_i$, $s_i$ determination since each of the free parameters 
is constrained by many bins. 

\begin{figure}
  \epsfig{file=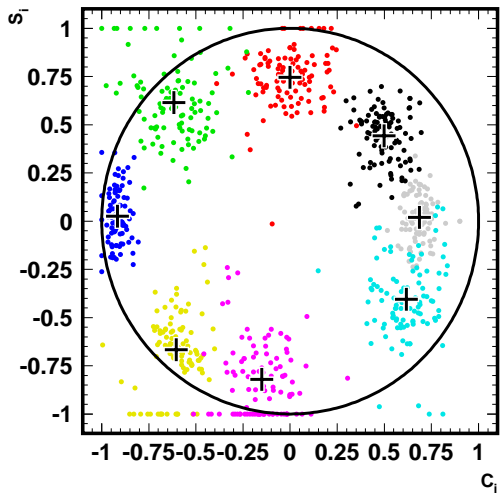, width=0.4\textwidth}
  \vspace{-\baselineskip}
  \caption{Coefficients $c_i$, $s_i$ calculated from  
           2000 generated $(K^0\pi\pi)^2$ events (points). 
           $\Delta\delta_D$ binning with $\mathcal{N}=8$, 
           100 toy MC samples are used. Different colors 
           correspond to different bins. Crosses denote the true 
           $c_i, s_i$ values in each bin. 
           }
  \label{cspic}
\end{figure}

The toy MC simulation was performed to study the procedure described above. 
Using the amplitude from the Belle analysis, we generate a large number 
of $D^0\to K^0_S\pi^+\pi^-$
decays and several sets of $(K^0_S\pi\pi)^2$ decays 
(according to the probability density given by (\ref{p_corr})). We use the 
same binning options as in $D_{CP}$ study with $\mathcal{N}=8$. 
The negative logarithmic likelihood (\ref{corr_lh}) 
is then minimized with $c_i$, $s_i$ and $h_{\rm corr}$ as free parameters. 
We constrain $|c_i,s_i|<1$ in the fit to improve the convergence. The 
coefficients $c_i$, $s_i$ are then used in the fit to $B$ decay data to obtain 
the $x,y$ error due to $(K^0\pi\pi)^2$ decay statistics. The number of $(K^0_S\pi\pi)^2$
decays ranges from 1000 to 10000. The obtained error shows a square root scaling. 
The best $(K^0\pi\pi)$-statistical error is obtained for $\Delta\delta_D$-binning
and recalculated to 1000 events yields $\sigma_x=0.0050$, $\sigma_y=0.0095$, which is 
only slightly worse than the error obtained with the same amount of $D_{CP}$ data
(see Table~\ref{stat} for comparison). 
We also check that changing the model used to define the binning does not 
lead to the systematic bias (although it does decrease the statistical 
precision). Figure~\ref{cspic} demonstrates the precision of the 
determination of $c_i$, $s_i$ coefficients in our toy MC study and the 
absence of the systematic bias for both $c_i$ and $s_i$. 

Since the number of $(K^0\pi\pi)^2$ decays in $\psi(3770)$ data is approximately 
twice larger than the number of $D_{CP}$ decays, the statistical errors due to 
$\psi(3770)$ data for the two approaches are nearly equal. The same binning 
can be used in both approaches, therefore improving the accuracy of $c_i$ determination. 
The approach based on $(K^0\pi\pi)^2$ data allows to extract both $c_i$ and $s_i$ 
without additional model uncertainties, so it can be used to check the validity 
of the constraint $c^2_i+s^2_i=1$ and therefore to test the sensitivity of the 
particular binning. 

\section{Conclusion}

We have studied the model-independent approach to $\phi_3$ measurement 
using $B^{\pm}\to DK^{\pm}$ decays 
with neutral $D$ decaying to $K^0_S\pi^+\pi^-$. 
The analysis of $\psi(3770)\to D\bar{D}$ data allows to extract
the information about the strong phase in \dkpp\ decay
that is fixed by model assumptions in a model-dependent technique.
We specially consider the case with a limited $\psi(3770)\to D\bar{D}$ data 
sample which will be available from CLEO-c in the nearest future. 

In the binned analysis, we propose a way to obtain the binning that 
offers an optimal statistical precision (close to the precision of 
an unbinned approach). 
Two different strategies of the binned analysis are considered: 
using $D_{CP}\to K^0_S\pi^+\pi^-$ data sample, and using decays of 
$\psi(3770)$ to $(K^0\pi^+\pi^-)_D (K^0\pi^+\pi^-)_D$. 
The strategy using $D_{CP}$ decays alone cannot offer a completely 
model-independent measurement: it provides only the information 
about $c_i$ coefficients, while $s_i$ for low $D_{CP}$ statistics 
has to be fixed using model assumptions. However, as the $D_{CP}$
data sample increases, model-independence can be reached by reducing 
the bin size. The strategy using the 
$\psi(3770)\to(K^0\pi^+\pi^-)_D (K^0\pi^+\pi^-)_D$ sample, in contrast, 
allows to obtain both $c_i$ and $s_i$ with an accuracy comparable to 
$D_{CP}$ approach. Both strategies can use the same binning of the \dkpp\ 
Dalitz plot and therefore can be used in combination to improve the accuracy 
due to $\psi(3770)$ statistics. 

The expected sensitivity is obtained based on the 
$D^0$ decay model from Belle analysis. For the CLEO-c statistics of 750 pb$^{-1}$ 
(1000 $D_{CP}$ events and 2000 $(K^0\pi\pi)^2$ events) the expected errors 
of parameters $x$ and $y$ due to $\psi(3770)$ statistics are 
$\sigma_x = 0.003$ and $\sigma_y = 0.007$. For $r_B=0.1$ it gives 
the $\phi_3$ precision 
$\sigma_{\phi_3}=\max(\sigma_x,\sigma_y)/(\sqrt{2}r_B)\simeq 3^{\circ}$, 
which is far below the expected error due to present-day $B$ data sample. 

In our study, we did not consider the experimental systematic uncertainties 
{\em e.g.} due to imperfect knowledge of the detection efficiency or 
background composition. We believe these issues can be addressed in a 
similar manner as in already completed model-dependent analyses. 

\section{Acknowledgments}

We would like to thank David Asner, Tim Gershon, Jure Zupan and 
Simon Eidelman for fruitful discussions and sugestions on improving 
the paper.

\end{document}